\documentclass[prx,aps,twocolumn,superscriptaddress,balancelastpage]{revtex4-1}
\usepackage{latexsym,amssymb,bm,bbold,amsfonts,amsmath,amstext,graphicx,bbm,bm,relsize,dsfont,theorem,times,xr}
\usepackage[dvipsnames]{xcolor}

\usepackage[unicode=true, breaklinks=false, pdfborder={0 0 1}, backref=false, colorlinks=true, linkcolor=blue, citecolor=blue]{hyperref}

\usepackage{subfiles}
 
\newcommand{\tr}{\mbox{tr}}

\newcommand{\rem}[1]{\textbf{\color{red}[[#1]]}}
\newcommand{\add}[1]{{\color{blue}#1}}
\newcommand{\com}[1]{\textbf{\color{OliveGreen}[[#1]]}}
\newcommand{\qq}[1]{\textbf{\color{RoyalBlue}#1}}
\def\tr{\mbox{tr}}
\def\bra#1{\langle{#1}|}
\def\ket#1{|{#1}\rangle}
\def\braket#1{\langle{#1}\rangle}
\def\Bra#1{\left\langle#1\right|}
\def\Ket#1{\left|#1\right \rangle}
\def\BraVert{\egroup\,\mid\,\bgroup}
\def\Brak#1#2#3{\bra{#1}#2\ket{#3}}
\def\ep{\epsilon}

\newcommand{\f}{\mathbf{F}}
\newcommand{\s}{\mathbf{S}}
\newcommand{\w}{\mathbf{W}}
\newcommand{\q}{\mathbf{Q}}
\newcommand{\x}{\mathbf{X}}

\newtheorem{definition}{Definition}
\newtheorem{proposition}[definition]{Proposition}

\newtheorem{theorem}[definition]{Theorem}

\newtheorem{conjecture}[definition]{Conjecture}

\begin{document}

\title{Enhancing the charging power of quantum batteries}

\author{Francesco Campaioli}
\email{francesco.campaioli@monash.edu}
\affiliation{School of Physics and Astronomy, Monash University, Victoria 3800, Australia}

\author{Felix A. Pollock}
\affiliation{School of Physics and Astronomy, Monash University, Victoria 3800, Australia}

\author{Felix C. Binder}
\affiliation{School of Physical \& Mathematical Sciences, Nanyang Technological University, 637371 Singapore, Singapore}

\author{Lucas~C{\'e}leri}
\affiliation{Instituto de F{\'i}sica, Universidade Federal de Goi{\'a}s, Caixa Postal 131, 74001-970, Goi{\^a}nia, Brazil}

\author{John Goold}
\affiliation{The Abdus Salam International Centre for Theoretical Physics (ICTP), Trieste, Italy}

\author{Sai Vinjanampathy}
\affiliation{Department of Physics, Indian Institute of Technology Bombay, Mumbai 400076, India} 
\affiliation{Centre for Quantum Technologies, National University of Singapore, 3 Science Drive 2, 117543 Singapore, Singapore} 

\author{Kavan Modi}
\email{kavan.modi@monash.edu}
\affiliation{School of Physics and Astronomy, Monash University, Victoria 3800, Australia}

\date{\today}

\begin{abstract}
Can collective quantum effects make a difference in a meaningful thermodynamic operation? Focusing on energy storage and batteries, we demonstrate that quantum mechanics can lead to an enhancement in the amount of work deposited per unit time, \textit{i.e.}, the charging power, when $N$ batteries are charged collectively. We first derive analytic upper bounds for the collective \emph{quantum advantage} in charging power for two choices of constraints on the charging Hamiltonian. We then demonstrate that even in the absence of quantum entanglement this advantage can be extensive. For our main result, we provide an upper bound to the achievable quantum advantage when the interaction order is restricted, \textit{i.e.}, at most $k$ batteries are interacting. This constitutes a fundamental limit on the advantage offered by quantum technologies over their classical counterparts.
\vspace{5pt}
\newline
DOI: \href{https://doi.org/10.1103/PhysRevLett.118.150601}{10.1103/PhysRevLett.118.150601}
\end{abstract} 
\maketitle

\makeatletter

{\bf Introduction --} Technology is currently being miniaturised at such a rate that we must give serious thought to the fundamental laws and blueprints of the machines of the future. In the microscopic domain, where these machines are expected to function, fluctuations of both thermal and quantum nature begin to proliferate, and quantum effects must be included in any reasonable physical description. When we deal with technologies working in this quantum regime, familiar thermodynamic concepts like work, heat, and entropy need to be applied with great care and consideration. It comes as no surprise that there has been a recent intense effort to understand how the laws of thermodynamics generalise to arbitrary quantum systems away from equilibrium. This effort is known as quantum thermodynamics and, given current interest in the development of quantum technologies, it is receiving a great deal of attention across a wide range of scientific communities~\cite{Goold:16, Millen:16, Vinjanampathy:15}.

Despite current momentum in the field of quantum thermodynamics, the explicit role of genuinely quantum features in the operation of thermal machines is not fully understood. A common issue raised is that the universal applicability of thermodynamics is rooted in the theory's complete lack of respect for microscopic details. So why then should thermodynamics really care about quantum mechanics? For example, the striking feature of Carnot's bound for the efficiency of a heat engine lies in the fact that it is insensitive to microscopic details~\cite{Gardas}. Nevertheless, if one relaxes the assumptions of large system size and quasi-static conditions, it is absolutely reasonable to get corrections based on the fine details of the working medium~\cite{Bender2000, Scully2003, Brunner2014, Uzdin2015}. An important question is then: can such quantum features be harnessed to improve other thermodynamically meaningful figures of merit, such as power?

Collective quantum phenomena are known to offer advantages in areas such as computation, secure communication, and metrology. Very recently, these possible advantages have received some attention in the context of batteries~\cite{Alicki:13, Hovhannisyan:13, Giorgi2015, Bruschi:15, Friis:15, Huber:15, Binder15b, Perarnau-Llobet:15,  BinderThesis}. The issue is subtle, in particular when one deals with mixed states. In particular, Alicki and Fannes suggested that entangling operations lead to increased work extraction from an energy storage device which they coined a ``quantum battery''~\cite{Alicki:13}. Nonetheless, while entangling operations are necessary for optimal work extraction, it has been shown that protocols exist for which no entanglement is actually created during optimal work extraction~\cite{Hovhannisyan:13, Giorgi2015}. Furthermore, considering a regime where entangling operations do not increase the extractable energy, some of the authors of the present work recently showed that entangling operations can, nonetheless, enhance the charging power of collections of two level quantum batteries~\cite{Binder15b}, also see~\cite{GelbwaserKlimovsky2015329}. However, the demonstration was reliant on a highly nonlocal Hamiltonian, which may be difficult to implement in practice.

In this Letter, we first formally define the collective {\it quantum advantage} for thermodynamic power, before deriving its ultimate upper bound. Next, we show that attaining a quantum advantage requires entangling operations, but not entanglement itself. We then go on to analytically prove that, for charging fields with finite interaction order, \textit{i.e.}, involving at most $k$-body interaction terms, the quantum advantage is upper bounded by a quadratic function of $k$ and cannot scale with the total number of batteries. Our result is a fundamental limit on how large power can be for physically realisable charging schemes, where the achievable interaction order is typically constrained.

{\bf Quantum Batteries --} We begin by defining what we mean by a quantum battery: Consider a quantum system with an internal Hamiltonian $I$. Such a system can be used to store work by manipulating an external control field $V(t)$ over some time interval $t \in (0,T)$. This generates the unitary dynamics $U =\vec{\mathcal{T}} \exp\{-i\int_{0}^{T} {\rm d}t H(t)\}$, where $H(t) = I+V(t)$ and $\vec{\mathcal{T}}$ is the time ordering operator (we set $\hbar=1$). Note that $V(t)$ vanishes outside the time interval $(0,T)$. During the charging process, the system is taken from an initial state $\rho$  to a higher energy final state $\sigma=U\rho \, U^\dag$ in a time $T$. Since the evolution is unitary, there is no heat generated~\cite{Alicki1979}, and the work deposited onto the system is given by $W = \tr[I (\sigma - \rho)]$ with an average charging power given by $P=W/T$.

Now, consider $N$ such batteries, whose joint initial state is $\rho^{\otimes N}$. As before, we can deposit work on all of them by transforming $\rho^{\otimes N}$ into $\sigma^{\otimes N}$. One way to implement this transformation is to perform the charging in parallel, following exactly the procedure described above for each battery independently. In this case, the unitary transformation is simply $U^{\otimes N}$, and the time taken to charge $N$ batteries is equal to the single battery charging time: $T_{\|} = T$. Since the deposited work scales extensively, $W_{\|}=N W$, leading to a charging power $P_{\|} = N P$ that grows linearly with the number of batteries.

Alternatively, to deposit work onto an array of $N$ batteries we could apply a more general unitary transformation $\bm{U}$, generated by the time-dependent $N$-battery Hamiltonian $\bm{H}(t) = \sum_{j=1}^N I^{(j)} + \bm{V}(t)$. Here, $I^{(j)}$ is the internal Hamiltonian for the $j$-th battery, and we require that $\bm{V}(t)$ vanishes outside the time interval $(0,T_\sharp)$. A bold font here denotes many-body operators. Henceforth, time-dependence will be left implicit where it is unambiguous. The crucial difference between $U^{\otimes N}$ and $\bm{U}$ is that $\bm{V}$ may contain terms corresponding to interactions between batteries -- \textit{i.e.}, the batteries are charged collectively. In this case the collective state of $N$-batteries $\bm{\rho}$ may become entangled. As before, we require that the system is transformed from state $\rho^{\otimes N}$ to state $\sigma^{\otimes N}$ via a cyclical operation; this ensures that the deposited work is the same as in the parallel case, \textit{i.e.}, $W_{\sharp} = N W = W_{\|}$. However, the joint time-evolved state $\bm{\rho}$ in this case may be entangled, and the time taken to implement $\bm{U}$ and $U$ could in principle be different, with $T_\sharp \le T_{\|}$ in the optimal case. This in turn leads to different charging powers: $P_\sharp \ge P_{\|}$.

{\bf Quantum advantage --} We are now in a position to define the \textit{quantum advantage} for collective charging as
\begin{gather}
\label{qadvantage}
\Gamma:=\frac{P_\sharp}{P_{\|}} = \frac{T_{\|}}{T_\sharp}, 
\end{gather}
where the second equality is a consequence of our requirement that the work done is independent of the charging method. Here, ``quantum'' refers specifically to an enhancement over charging with the best local (\textit{i.e.}, non-entangling) operations. That is, to compute the quantum advantage we must take the optimal values for $P_{\|}$ ($P_{\sharp}$) for given $\rho$ ($\rho^{\otimes N}$) and $\sigma$ ($\sigma^{\otimes N}$).

In order for this advantage to be meaningful, we must ensure that the parallel and collective charging strategies are fairly compared. In particular, we would like to isolate the advantage due to collective quantum effects without worrying about other consequences of introducing interactions between batteries, such as the increased energy available to drive transitions. In order to take this extra energy into account, we must constrain the collective Hamiltonian $\bm{H}$ to be similar to $\sum_j H^{(j)}$, the total Hamiltonian in the parallel charging case. Without constraints, we could freely increase the total energy of the collective charging Hamiltonian to achieve faster driving, making the advantage arbitrarily large. Noting that the variance and mean energy are extensive quantities for non-interacting systems, we consider two possible constraints on $\bm{H}$, namely:

\vspace{5pt}
\noindent\textbf{(C1)} The time-averaged standard deviation in energy during the collective evolution for time $T_\sharp$ should not exceed $\sqrt{N}$ times that of a single battery, \textit{i.e.}, $\Delta E_\sharp \le \sqrt{N} \Delta E $ with 
\begin{align}
\label{con:var}
& \Delta E_\sharp  := \int_0^{T_\sharp} \! dt \,
\frac{\Delta \bm{H}_{\bm{\rho}}}{T_\sharp}, \quad
\Delta E := \int_0^{T_\|} \! dt \,
\frac{\Delta H_{\rho}}{T_\|},
\end{align}
where $(\Delta X_y)^2 := \braket{(X-\braket{X}_y)^2}_y$ and $\braket{X}_y=\tr[X y]$.
\vspace{.2cm}

\noindent\textbf{(C2)} The time-averaged energy during the collective evolution for time $T_\sharp$ should not exceed $N$ times that of a single battery, \textit{i.e.}, $E_\sharp \le N E $ with
\begin{align}\label{con:en}
& E_\sharp := \int_0^{T_\sharp} \! dt \, 
\frac{\braket{\bm{H}-\bm{h}_g}_{\bm{\rho}}}{T_\sharp}
, \quad
E := \int_0^{T_\|} \! dt \, 
\frac{\braket{H-h_g}_\rho}{T_\|},
\end{align}
where $\bm{h}_g$ ($h_g$) is the instantaneous ground state energy of $\bm{H}$ ($H$), that can also depend on time.

We are free to choose either of these rescalings as a constraint on $\bm{H}$. They represent alternative ways of accounting for the differing energetic structure of the parallel and collective charging Hamiltonians. The choices of C1 and C2 are additionally motivated by the form of the quantum speed limit (see below). While C1 leads to a stricter upper bound on the quantum advantage, there is no reason \emph{a priori} to choose one over the other. We are now ready to derive our first main result.

{\bf Upper bound --} Since the quantum advantage defined in Eq.~\eqref{qadvantage} amounts to a ratio of transition times, we can use the quantum speed limit (QSL) to upper bound it for a given constraint. The QSL states that the time required to transform $\rho^{\otimes N}$ into $\sigma^{\otimes N}$ is lower bounded as $T_\sharp\geq T^{(N)}_{\rm QSL} := \mathcal{L}_{N} \max\left(1/E_\sharp, 1/\Delta E_\sharp\right)$, where $\mathcal{L}_m:= \arccos(\sqrt{F[\rho^{\otimes m}, \sigma^{\otimes m}}])$ is the Bures angle and $F(\rho,\sigma):= \tr[\sqrt{\sqrt{\rho} \, \sigma \sqrt{\rho}}]^{2}$ is the Uhlmann fidelity~\cite{Deffner:14}. The two constraints, C1 and C2, are clearly related to the QSL, as $\Delta E_\sharp$ and $E_\sharp$ can be computed using Eqs.~\eqref{con:var} and \eqref{con:en} respectively. If we could change the Hamiltonian at will to include arbitrary interaction terms, and only concern ourselves with the state transformation, we could replace $\bm{H}$ with a time-independent Hamiltonian such that the initial state traverses the great circle connecting $\rho^{\otimes N}$ and $\sigma^{\otimes N}$ in the $N$-partite state space~\cite{PhysRevLett.110.157207, PhysRevLett.111.260501, PhysRevA.90.032110}. In the absence of further constraints, the evolution with such time-independent Hamiltonians is in fact optimal. In this case, the QSL reduces to the usual inequalities due to Mandelstam-Tamm~\cite{Mandelstam:45} and Margolus-Levitin~\cite{Margolus:98}, where $E_\sharp$ is replaced by the average initial energy and $\Delta E_\sharp$ is replaced by the average initial standard deviation.

To derive the upper bound, we first confine ourselves to constraint C1. We proceed by noting that the quantum speed limit for collectively charging $N$ quantum batteries is given by $T_\sharp \geq \mathcal{L}_{N}/\Delta E_\sharp$. This means that $\Gamma \leq T_{\|} \Delta E_\sharp / \mathcal{L}_{N}$. Now, using constraint C1, \textit{i.e.}, $\Delta E_\sharp \le \sqrt{N} \Delta E$, we get $\Gamma \leq T_\| \sqrt{N} \Delta E/\mathcal{L}_N$. A similar argument can be made with constraint (C2). Taking into account that the speed limit for parallel charging is not always attainable, we arrive at the following upper bounds for the quantum advantage:
\begin{gather}\label{conservativebounds}
\Gamma_{\rm C1} \le \beta \sqrt{N} \frac{\mathcal{L}_{1}}{\mathcal{L}_{N}} 
\quad \mbox{and} \quad
\Gamma_{\rm C2} \le \beta N \frac{\mathcal{L}_{1}}{\mathcal{L}_{N}},
\end{gather}
for constraints \textbf{(C1)} and \textbf{(C2)} respectively, where $\beta:=T_{\rm \|}/T^{(1)}_{\rm QSL}$ quantifies the inability to saturate the QSL in the parallel case.

Two remarks are in order: Firstly, for orthogonal pure initial and final states, the QSL can be saturated and $\beta=1$. Though the quantum advantage for power could be larger in other cases, including where the battery states are mixed~\cite{Pires2016}, the improvement cannot grow with the number of batteries; \textit{i.e.}, $\beta$ is a constant function of $N$. Secondly, we have excluded cases where $\rho$ and $\sigma$ do not lie on the same unitary orbit, as there is no way of transforming the former into the latter using the scheme outlined above; the two states will therefore necessarily have the same spectrum~\cite{Binder15a}.

The two bounds in Eq.~\eqref{conservativebounds} are independent from each other, and constraint C1 is stronger than C2, as it leads to a stricter bound on the quantum advantage. Many other bounds can be derived by considering other extensive constraints. The quantum advantage is tight for orthogonal initial and final states, due to the example given in Ref.~\cite{Binder15b}.

The significance of entanglement for quantum enhancement has previously been studied in the context of quantum speed limits for pure states: it was shown that, for non-interacting systems, initial entanglement is required for an enhancement in the speed of evolution~\cite{Giovannetti2003, Zander2007}, while for interacting  systems a speedup may be achieved for initially separable states, since intermediate entangled states are accessible~\cite{Giovannetti2003a, Giovannetti2003b, Xu2015}. In the more general case of mixed states, the necessity of entanglement for an enhancement may not be directly inferred, though it has been argued that, in general, larger quantum Fisher information of the state with respect to the generator of evolution leads to enhanced speed~\cite{Frowis2012, Toth2014}. In fact, as we now show, entanglement does not appear to be necessary for a nontrivial quantum advantage.

\begin{proposition}
\label{thm:sepball}
An extensive quantum advantage can be attained even for highly mixed states, including those confined to the separable ball throughout the charging procedure.
\end{proposition}

We prove this with an explicit example: Consider $N$ two-level batteries with internal Hamiltonian $I$ with eigenstates $\ket{E_1}$ and $\ket{E_0}$, and corresponding energies $E_1=1$ and $E_0=0$. Let the initial state be thermal: $\rho=\exp(-\epsilon I)/\mathcal{Z}$ at inverse temperature $\epsilon$ with $\mathcal{Z}=\tr[\exp(-\epsilon I)]$, and the final state be $\sigma=\exp(\epsilon I)/\mathcal{Z}$. The optimal local charging scheme is achieved in time $T_\|=\pi/2$ by applying Hamiltonian $H = \ket{E_0}\bra{E_1} + \ket{E_1}\bra{E_0}$ to each battery. In contrast, the joint charging of $N$ batteries is achieved in $T_\sharp = T_{\|} / \alpha_\sharp$ using the \emph{global} Hamiltonian $\bm{H}_\sharp = \alpha_\sharp H^{\otimes N}$, where the positive constant $\alpha_\sharp$ is introduced to satisfy the chosen constraint.

In both cases (local and global) the deposited work is identical; thus, the quantum advantage is simply the ratio of $T_\|$ to $T_\sharp$: $\Gamma = \alpha_\sharp$, which can be evaluated for the choice of constraint. We find $\Gamma_{\textrm{C1}} = \sqrt{N}$ and $\Gamma_{\textrm{C2}} = N$ (also $\Gamma_{\textrm{C0}}=N$ for C0 given in Eq.~\eqref{con:opnorm}).

For $N$ quantum systems of d-dimensions, there exists a ball of radius $R(N,d)$, centred on the maximally mixed state, containing only separable states~\footnote{Though an exact form for the radius of the separable ball is not known, it has been bounded from below and above~\cite{GurvitsBarnum2002, GurvitsBarnum2005, AubrunSzarek2006}.}. Since the distance from the maximally mixed state cannot change under unitary evolution, for a small enough choice of $\epsilon$, the joint state of $N$ batteries will lie within this ball throughout the evolution; yet, the quantum advantage remains extensive. \hfill $\blacksquare$

Remarkably, neither $T_\|$ nor $T_\sharp$ depend on $\epsilon$, while the total work done does. In other words, no matter how mixed the battery is a quantum advantage that scales with the number of batteries involved is always achievable. The trade-off of using highly mixed states is that the charging power suffers as $\epsilon$ becomes smaller and smaller. Proposition~\ref{thm:sepball} implies that, while a quantum advantage requires entangling operations, the joint state of $N$-batteries does not have to be entangled during the charging process.

The Hamiltonian used in the example above, and in Ref.~\cite{Binder15b}, to saturate the bound for quantum advantage involves $N$-body interactions. Such interactions are notoriously difficult to engineer. In the next section, we consider physically realisable interactions, and study the dependence of the enhancement on the order of the charging interaction, \textit{i.e.}, the number of batteries that take part in a single interaction term.

{\bf $k$-local charging --}
We now investigate the achievability of a significant quantum advantage in a regime where arbitrary multipartite entanglement generation is possible during the charging process. In particular, we demonstrate that, although a nontrivial quantum advantage is achievable in physical systems characterised by at most $k$-body interactions, this advantage -- upper bounded by a quantity that depends at most quadratically on $k$ -- cannot scale with the number $N$ of batteries that compose the system.

First, we consider the situation where work is deposited onto the battery by means of a piecewise unitary circuit, an example of which is depicted in Fig.~\ref{fig:pairwise_circuit} in the supplementary material (SM). This model is reminiscent of the circuit model of universal quantum computation, which is known to outperform its classical counterpart. In this case, the collective state of N-batteries will, in general, be highly entangled. This scheme allows us to study how the quantum advantage is related to the number of batteries that are simultaneously interacting.

We consider batteries composed of $N$  $d$-level systems, with internal Hamiltonian $\bm{I}=\sum_{j} I^{(j)}$, as before. More explicitly, and without loss of generality, we assume that each term is given by $I^{(j)}=\sum_{l=1}^d \lambda_l \ket{l}_j\bra{l}_j$ with $\lambda_d-\lambda_1 = 2\lambda_d >0$, and with eigenvalues arranged in increasing order. The time interval $[0,T_\sharp]$ is divided up into $L$ steps: at each step the Hamiltonian is the sum of $s=\lceil N/k \rceil$ terms, each acting on a different set of $k$ batteries. In order to allow the formation of highly entangled states, these partitions could be different at each step.  At any time $t$, the $k$-local Hamiltonian can be written as $\bm{H}=\sum_{\mu=1}^s h_{\mu}$ where each term $h_{\mu}$ acts on a different $k$-partition of the Hilbert space, identified by the set $\mu=(\mu_1,\dots,\mu_k)$ of $k$ indices.

In order to make a meaningful statement in this scenario, we need to introduce a third constraint:

\vspace{5pt}
\noindent\textbf{(C0)} The time-averaged operator norm of the driving Hamiltonian $\bm{H}$ during the collective evolution for time $T_\sharp$ should not exceed $N$ times that of a single battery driving Hamiltonian, \textit{i.e.}, $\mathcal{E}_\sharp \leq N \mathcal{E}$ with 
\begin{gather}\label{con:opnorm}
 \mathcal{E}_\sharp  := \frac{1}{T_\sharp}\int_0^{T_\sharp} dt \, \lVert \bm{H} \rVert_{\rm op} \;\, \mbox{and} \;\,
 \mathcal{E} := \frac{1}{T_\|}\int_0^{T_\|} dt \, \lVert {H} \rVert_{\rm op},
\end{gather}
where the operator norm $\|A\|_{\rm op}$ is defined as the largest singular value of $A$. 

Constraint C0 guarantees that both the time-averaged standard deviation and the time-averaged energy are bounded from above, as shown in Sec.~\ref{s:bounds} of SM. There, we show that $\mathcal{E}_\sharp$ upper bounds both $E_\sharp/2$ and $\Delta E_\sharp$. In this sense, it is a stricter constraint than C1 or C2.

We now show that, with this constraint, the upper bound on the quantum advantage depends on the interaction order $k$:

\begin{theorem}\label{thm:circuit}
For a circuit based charging procedure with interaction order of at most $k$, the achievable quantum advantage is upper bounded as $\Gamma_{\rm C0} < \gamma  k$, where $\gamma$ is a constant that does not scale with the number $N$ of batteries.
\end{theorem}

Proof in Sec.~\ref{s:th2} of SM. In the important case where $\rho$ and $\sigma$ are the ground and maximally excited states respectively, $\gamma=\pi/2$. By construction, the bound on the quantum advantage is not tight. For comparison, it has been shown elsewhere that $\Gamma_{\rm C1} = \sqrt{k}$ and $\Gamma_{\rm C2} = k$ are achievable if the total number of batteries $N$ can be divided by $k$, \textit{i.e.}, if $N/k=s\in \mathbb{N}$~\cite{Giovannetti2004,Binder15b}. In this particular case, such a speed-up can be obtained for pure states using the time-independent Hamiltonian $\bm{H} = \sqrt{s} \sum_{\mu=1}^s h_\mu$, with $h_\mu = |1\rangle^{\otimes k} \langle d|^{\otimes k} + h.c.$, assuming that each $h_\mu$ acts on a completely different set of $k$ batteries, \textit{i.e.}, $[h_\mu,h_{\mu'}]=0$  $\,\forall \mu, \mu'$. In the same situation, using constraint C0 we obtain $\Gamma_{\rm C0}= k$, suggesting that the strict inequality in Theorem~\ref{thm:circuit} is only different by a constant factor from an achievable bound.

Theorem~\ref{thm:circuit} can be extended to more general cases, where $k$-body time-dependent interactions can occur between overlapping sets of batteries, with the restriction that each battery is simultaneously interacting with at most $m$ others. This restriction is motivated by the idea that the \emph{reach} of the interaction should be limited.

\begin{theorem} \label{thm:reach}
For a generic time-dependent charging procedure, the achievable quantum advantage is upper bounded as $\Gamma_{\rm C0} < \gamma \big(k^2(m-1) + k\big)$,
where $k$ is the interaction order and $m$ is the maximum participation number.
\end{theorem}

The proof is given in Sec.~\ref{s:co1} of SM. For many physical systems, both $k$ and $m$ are limited: 2 or 3-body interactions are the norm for fundamental processes, and higher interaction orders are generally hard to engineer here~\cite{zoller, Boxio:07, napolitano2011interaction}.
The effective participation number, or reach, $m$ tends to be constrained by the spatial arrangement of systems and the fact that interaction strength often drops off with distance. Exceptions to this include the Dicke model~\cite{PhysRev.93.99} where collective coherence leads to superradiance, the Lipkin-Meshkov-Glick model~\cite{LIPKIN1965188}, where all particles interact with each other, and  the M{\o}lmer-S{\o}rensen interaction~\cite{MolmerSorensen1998}, in which an ensemble of ions are effectively coupled by a spatially uniform electromagnetic field.

Note that these bounds are not tight; while a scaling of the power $P_\sharp$ with the number of batteries $N$ is surely not feasible in the context of $k$-body interactions, it is more likely that the quantum advantage is tightly limited by $k$. In fact, we conjecture that, for any choice of $\bm{H}$, a conservative bound for the quantum advantage is given by $\Gamma_{\rm C0} < \gamma k$:

\begin{conjecture}\label{conj:conj}
Theorem 2 holds for any time-dependent $k$-body interaction Hamiltonian.
\end{conjecture}

We examine this particular statement in SM Sec.~\ref{s:conjecture}, anticipating that the result holds if a particular mathematical conjecture does too. While we cannot exclude measure zero cases, a large sample of charging Hamiltonians (generated from Haar-random unitary operations), with $(N,k)=(3,2)$, $(4,2)$, $(4,3)$ and $(6,2)$, has failed to produce any counterexamples. We believe that similar conjectures should also hold for constraints C1 and C2.

{\bf Conclusions --} In this Letter, we have introduced the notion of collective quantum advantage for thermodynamic power. Our results directly complement a previous strain of research into quantum speed limits, by deriving a concrete upper bound on the ratio between the maximum speed of interacting and non-interacting driving between separable states. We have proven two fundamental upper bounds for the quantum advantage, each corresponding to a different constraint on the charging Hamiltonian. We have also shown analytically that a quantum advantage that grows with the number of batteries is not achievable with any physically reasonable Hamiltonian (\textit{i.e.}, one with at most $k$-body interactions). Nevertheless, a quantum advantage that grows with the interaction order $k$ can be achieved.

The quantum advantage has been interpreted as the result of rapid evolution through the space of high-dimensional quantum states, typically obtained by means of global operations~\cite{Binder15b}. While, in the case of pure states, entanglement is a necessary consequence of these global operations, a fully separable evolution is still accessible for those states that live in the separable ball.

A striking consequence of our results, which hold in general for mixed states, is that an enhanced charging power is available even for arbitrarily mixed states, in remarkable analogy to the case of quantum metrology. There, an enhancement in sensing is still available for highly mixed states lying inside the separable ball~\cite{prx}.

While collective behaviour has been demonstrated to provide an advantage in performing many information theoretic tasks, such distinctions from classical behaviour are few and far between in thermodynamics. This Letter demonstrates that thermodynamic processes can indeed benefit from collective effects when time enters the picture, though physical limitations on the interaction order prevent us from utilising them. This result has fundamental importance for our understanding of how quantum theory and thermodynamics are related.

\begin{acknowledgments}
The authors would like to thank an anonymous referee for helpful comments. Centre for Quantum Technologies is a Research Centre of Excellence funded by the Ministry of Education and the National Research Foundation of Singapore. This work was partially supported by the COST Action MP1209. FB acknowledges support by the National Research Foundation of Singapore (Fellowship NRF-NRFF2016-02). KM and LCC acknowledge support by CNPq (Grants No. CNPq: 401230/2014-7). LCC acknowledges further support by CNPq (Grants No. CNPq: 305086/2013-8 and 445516/2014-3).
\end{acknowledgments}


\bibliography{battery}

\begin{thebibliography}{47}%
\makeatletter
\providecommand \@ifxundefined [1]{%
 \@ifx{#1\undefined}
}%
\providecommand \@ifnum [1]{%
 \ifnum #1\expandafter \@firstoftwo
 \else \expandafter \@secondoftwo
 \fi
}%
\providecommand \@ifx [1]{%
 \ifx #1\expandafter \@firstoftwo
 \else \expandafter \@secondoftwo
 \fi
}%
\providecommand \natexlab [1]{#1}%
\providecommand \enquote  [1]{``#1''}%
\providecommand \bibnamefont  [1]{#1}%
\providecommand \bibfnamefont [1]{#1}%
\providecommand \citenamefont [1]{#1}%
\providecommand \href@noop [0]{\@secondoftwo}%
\providecommand \href [0]{\begingroup \@sanitize@url \@href}%
\providecommand \@href[1]{\@@startlink{#1}\@@href}%
\providecommand \@@href[1]{\endgroup#1\@@endlink}%
\providecommand \@sanitize@url [0]{\catcode `\\12\catcode `\$12\catcode
  `\&12\catcode `\#12\catcode `\^12\catcode `\_12\catcode `\%12\relax}%
\providecommand \@@startlink[1]{}%
\providecommand \@@endlink[0]{}%
\providecommand \url  [0]{\begingroup\@sanitize@url \@url }%
\providecommand \@url [1]{\endgroup\@href {#1}{\urlprefix }}%
\providecommand \urlprefix  [0]{URL }%
\providecommand \Eprint [0]{\href }%
\providecommand \doibase [0]{http://dx.doi.org/}%
\providecommand \selectlanguage [0]{\@gobble}%
\providecommand \bibinfo  [0]{\@secondoftwo}%
\providecommand \bibfield  [0]{\@secondoftwo}%
\providecommand \translation [1]{[#1]}%
\providecommand \BibitemOpen [0]{}%
\providecommand \bibitemStop [0]{}%
\providecommand \bibitemNoStop [0]{.\EOS\space}%
\providecommand \EOS [0]{\spacefactor3000\relax}%
\providecommand \BibitemShut  [1]{\csname bibitem#1\endcsname}%
\let\auto@bib@innerbib\@empty
\bibitem [{\citenamefont {Goold}\ \emph {et~al.}(2016)\citenamefont {Goold},
  \citenamefont {Huber}, \citenamefont {Riera}, \citenamefont {del Rio},\ and\
  \citenamefont {Skrzypczyk}}]{Goold:16}%
  \BibitemOpen
  \bibfield  {author} {\bibinfo {author} {\bibfnamefont {J.}~\bibnamefont
  {Goold}}, \bibinfo {author} {\bibfnamefont {M.}~\bibnamefont {Huber}},
  \bibinfo {author} {\bibfnamefont {A.}~\bibnamefont {Riera}}, \bibinfo
  {author} {\bibfnamefont {L.}~\bibnamefont {del Rio}}, \ and\ \bibinfo
  {author} {\bibfnamefont {P.}~\bibnamefont {Skrzypczyk}},\ }\href
  {http://stacks.iop.org/1751-8121/49/i=14/a=143001} {\bibfield  {journal}
  {\bibinfo  {journal} {J. Phys. A: Math. Th.}\ }\textbf {\bibinfo {volume}
  {49}},\ \bibinfo {pages} {143001} (\bibinfo {year} {2016})}\BibitemShut
  {NoStop}%
\bibitem [{\citenamefont {Millen}\ and\ \citenamefont
  {Xuereb}(2016)}]{Millen:16}%
  \BibitemOpen
  \bibfield  {author} {\bibinfo {author} {\bibfnamefont {J.}~\bibnamefont
  {Millen}}\ and\ \bibinfo {author} {\bibfnamefont {A.}~\bibnamefont
  {Xuereb}},\ }\href {\doibase 10.1088/1367-2630/18/1/011002} {\bibfield
  {journal} {\bibinfo  {journal} {New J. Phys.}\ }\textbf {\bibinfo {volume}
  {18}},\ \bibinfo {pages} {011002} (\bibinfo {year} {2016})}\BibitemShut
  {NoStop}%
\bibitem [{\citenamefont {Vinjanampathy}\ and\ \citenamefont
  {Anders}(2016)}]{Vinjanampathy:15}%
  \BibitemOpen
  \bibfield  {author} {\bibinfo {author} {\bibfnamefont {S.}~\bibnamefont
  {Vinjanampathy}}\ and\ \bibinfo {author} {\bibfnamefont {J.}~\bibnamefont
  {Anders}},\ }\href {\doibase 10.1080/00107514.2016.1201896} {\bibfield
  {journal} {\bibinfo  {journal} {Contemp. Phys.}\ }\textbf {\bibinfo {volume}
  {57}},\ \bibinfo {pages} {545} (\bibinfo {year} {2016})}\BibitemShut
  {NoStop}%
\bibitem [{\citenamefont {Gardas}\ and\ \citenamefont
  {Deffner}(2015)}]{Gardas}%
  \BibitemOpen
  \bibfield  {author} {\bibinfo {author} {\bibfnamefont {B.}~\bibnamefont
  {Gardas}}\ and\ \bibinfo {author} {\bibfnamefont {S.}~\bibnamefont
  {Deffner}},\ }\href {\doibase 10.1103/PhysRevE.92.042126} {\bibfield
  {journal} {\bibinfo  {journal} {Phys. Rev. E}\ }\textbf {\bibinfo {volume}
  {92}},\ \bibinfo {pages} {042126} (\bibinfo {year} {2015})}\BibitemShut
  {NoStop}%
\bibitem [{\citenamefont {Bender}\ \emph {et~al.}(2000)\citenamefont {Bender},
  \citenamefont {Brody},\ and\ \citenamefont {Meister}}]{Bender2000}%
  \BibitemOpen
  \bibfield  {author} {\bibinfo {author} {\bibfnamefont {C.~M.}\ \bibnamefont
  {Bender}}, \bibinfo {author} {\bibfnamefont {D.~C.}\ \bibnamefont {Brody}}, \
  and\ \bibinfo {author} {\bibfnamefont {B.~K.}\ \bibnamefont {Meister}},\
  }\href {\doibase 10.1088/0305-4470/33/24/302} {\bibfield  {journal} {\bibinfo
   {journal} {J. Phys. A: Math. Gen.}\ }\textbf {\bibinfo {volume} {33}},\
  \bibinfo {pages} {4427} (\bibinfo {year} {2000})}\BibitemShut {NoStop}%
\bibitem [{\citenamefont {Scully}\ \emph {et~al.}(2003)\citenamefont {Scully},
  \citenamefont {Zubairy}, \citenamefont {Agarwal},\ and\ \citenamefont
  {Walther}}]{Scully2003}%
  \BibitemOpen
  \bibfield  {author} {\bibinfo {author} {\bibfnamefont {M.~O.}\ \bibnamefont
  {Scully}}, \bibinfo {author} {\bibfnamefont {M.~S.}\ \bibnamefont {Zubairy}},
  \bibinfo {author} {\bibfnamefont {G.~S.}\ \bibnamefont {Agarwal}}, \ and\
  \bibinfo {author} {\bibfnamefont {H.}~\bibnamefont {Walther}},\ }\href
  {http://science.sciencemag.org/content/299/5608/862} {\bibfield  {journal}
  {\bibinfo  {journal} {Science}\ }\textbf {\bibinfo {volume} {299}} (\bibinfo
  {year} {2003})}\BibitemShut {NoStop}%
\bibitem [{\citenamefont {Brunner}\ \emph {et~al.}(2014)\citenamefont
  {Brunner}, \citenamefont {Huber}, \citenamefont {Linden}, \citenamefont
  {Popescu}, \citenamefont {Silva},\ and\ \citenamefont
  {Skrzypczyk}}]{Brunner2014}%
  \BibitemOpen
  \bibfield  {author} {\bibinfo {author} {\bibfnamefont {N.}~\bibnamefont
  {Brunner}}, \bibinfo {author} {\bibfnamefont {M.}~\bibnamefont {Huber}},
  \bibinfo {author} {\bibfnamefont {N.}~\bibnamefont {Linden}}, \bibinfo
  {author} {\bibfnamefont {S.}~\bibnamefont {Popescu}}, \bibinfo {author}
  {\bibfnamefont {R.}~\bibnamefont {Silva}}, \ and\ \bibinfo {author}
  {\bibfnamefont {P.}~\bibnamefont {Skrzypczyk}},\ }\href {\doibase
  10.1103/PhysRevE.89.032115} {\bibfield  {journal} {\bibinfo  {journal} {Phys.
  Rev. E}\ }\textbf {\bibinfo {volume} {89}},\ \bibinfo {pages} {032115}
  (\bibinfo {year} {2014})}\BibitemShut {NoStop}%
\bibitem [{\citenamefont {Uzdin}\ \emph {et~al.}(2015)\citenamefont {Uzdin},
  \citenamefont {Levy},\ and\ \citenamefont {Kosloff}}]{Uzdin2015}%
  \BibitemOpen
  \bibfield  {author} {\bibinfo {author} {\bibfnamefont {R.}~\bibnamefont
  {Uzdin}}, \bibinfo {author} {\bibfnamefont {A.}~\bibnamefont {Levy}}, \ and\
  \bibinfo {author} {\bibfnamefont {R.}~\bibnamefont {Kosloff}},\ }\href
  {\doibase 10.1103/PhysRevX.5.031044} {\bibfield  {journal} {\bibinfo
  {journal} {Phys. Rev. X}\ }\textbf {\bibinfo {volume} {5}},\ \bibinfo {pages}
  {031044} (\bibinfo {year} {2015})}\BibitemShut {NoStop}%
\bibitem [{\citenamefont {Alicki}\ and\ \citenamefont
  {Fannes}(2013)}]{Alicki:13}%
  \BibitemOpen
  \bibfield  {author} {\bibinfo {author} {\bibfnamefont {R.}~\bibnamefont
  {Alicki}}\ and\ \bibinfo {author} {\bibfnamefont {M.}~\bibnamefont
  {Fannes}},\ }\href {\doibase 10.1103/PhysRevE.87.042123} {\bibfield
  {journal} {\bibinfo  {journal} {Phys. Rev. E}\ }\textbf {\bibinfo {volume}
  {87}},\ \bibinfo {pages} {042123} (\bibinfo {year} {2013})}\BibitemShut
  {NoStop}%
\bibitem [{\citenamefont {Hovhannisyan}\ \emph {et~al.}(2013)\citenamefont
  {Hovhannisyan}, \citenamefont {Perarnau-Llobet}, \citenamefont {Huber},\ and\
  \citenamefont {Ac\'{\i}n}}]{Hovhannisyan:13}%
  \BibitemOpen
  \bibfield  {author} {\bibinfo {author} {\bibfnamefont {K.~V.}\ \bibnamefont
  {Hovhannisyan}}, \bibinfo {author} {\bibfnamefont {M.}~\bibnamefont
  {Perarnau-Llobet}}, \bibinfo {author} {\bibfnamefont {M.}~\bibnamefont
  {Huber}}, \ and\ \bibinfo {author} {\bibfnamefont {A.}~\bibnamefont
  {Ac\'{\i}n}},\ }\href {\doibase 10.1103/PhysRevLett.111.240401} {\bibfield
  {journal} {\bibinfo  {journal} {Phys. Rev. Lett.}\ }\textbf {\bibinfo
  {volume} {111}},\ \bibinfo {pages} {240401} (\bibinfo {year}
  {2013})}\BibitemShut {NoStop}%
\bibitem [{\citenamefont {Giorgi}\ and\ \citenamefont
  {Campbell}(2015)}]{Giorgi2015}%
  \BibitemOpen
  \bibfield  {author} {\bibinfo {author} {\bibfnamefont {G.~L.}\ \bibnamefont
  {Giorgi}}\ and\ \bibinfo {author} {\bibfnamefont {S.}~\bibnamefont
  {Campbell}},\ }\href {\doibase 10.1088/0953-4075/48/3/035501} {\bibfield
  {journal} {\bibinfo  {journal} {J. Phys. B: At. Mol. Opt. Phys.}\ }\textbf
  {\bibinfo {volume} {48}},\ \bibinfo {pages} {035501} (\bibinfo {year}
  {2015})}\BibitemShut {NoStop}%
\bibitem [{\citenamefont {Bruschi}\ \emph {et~al.}(2015)\citenamefont
  {Bruschi}, \citenamefont {Perarnau-Llobet}, \citenamefont {Friis},
  \citenamefont {Hovhannisyan},\ and\ \citenamefont {Huber}}]{Bruschi:15}%
  \BibitemOpen
  \bibfield  {author} {\bibinfo {author} {\bibfnamefont {D.~E.}\ \bibnamefont
  {Bruschi}}, \bibinfo {author} {\bibfnamefont {M.}~\bibnamefont
  {Perarnau-Llobet}}, \bibinfo {author} {\bibfnamefont {N.}~\bibnamefont
  {Friis}}, \bibinfo {author} {\bibfnamefont {K.~V.}\ \bibnamefont
  {Hovhannisyan}}, \ and\ \bibinfo {author} {\bibfnamefont {M.}~\bibnamefont
  {Huber}},\ }\href {\doibase 10.1103/PhysRevE.91.032118} {\bibfield  {journal}
  {\bibinfo  {journal} {Phys. Rev. E}\ }\textbf {\bibinfo {volume} {91}},\
  \bibinfo {pages} {032118} (\bibinfo {year} {2015})}\BibitemShut {NoStop}%
\bibitem [{\citenamefont {Friis}\ \emph {et~al.}(2016)\citenamefont {Friis},
  \citenamefont {Huber},\ and\ \citenamefont {Perarnau-Llobet}}]{Friis:15}%
  \BibitemOpen
  \bibfield  {author} {\bibinfo {author} {\bibfnamefont {N.}~\bibnamefont
  {Friis}}, \bibinfo {author} {\bibfnamefont {M.}~\bibnamefont {Huber}}, \ and\
  \bibinfo {author} {\bibfnamefont {M.}~\bibnamefont {Perarnau-Llobet}},\
  }\href {\doibase 10.1103/PhysRevE.93.042135} {\bibfield  {journal} {\bibinfo
  {journal} {Phys. Rev. E}\ }\textbf {\bibinfo {volume} {93}},\ \bibinfo
  {pages} {042135} (\bibinfo {year} {2016})}\BibitemShut {NoStop}%
\bibitem [{\citenamefont {Huber}\ \emph {et~al.}(2015)\citenamefont {Huber},
  \citenamefont {Perarnau-Llobet}, \citenamefont {Hovhannisyan}, \citenamefont
  {Skrzypczyk}, \citenamefont {Kl{\"o}ckl}, \citenamefont {Brunner},\ and\
  \citenamefont {Ac{\'\i}n}}]{Huber:15}%
  \BibitemOpen
  \bibfield  {author} {\bibinfo {author} {\bibfnamefont {M.}~\bibnamefont
  {Huber}}, \bibinfo {author} {\bibfnamefont {M.}~\bibnamefont
  {Perarnau-Llobet}}, \bibinfo {author} {\bibfnamefont {K.~V.}\ \bibnamefont
  {Hovhannisyan}}, \bibinfo {author} {\bibfnamefont {P.}~\bibnamefont
  {Skrzypczyk}}, \bibinfo {author} {\bibfnamefont {C.}~\bibnamefont
  {Kl{\"o}ckl}}, \bibinfo {author} {\bibfnamefont {N.}~\bibnamefont {Brunner}},
  \ and\ \bibinfo {author} {\bibfnamefont {A.}~\bibnamefont {Ac{\'\i}n}},\
  }\href {\doibase 10.1088/1367-2630/17/6/065008} {\bibfield  {journal}
  {\bibinfo  {journal} {New J. Phys.}\ }\textbf {\bibinfo {volume} {17}},\
  \bibinfo {pages} {065008} (\bibinfo {year} {2015})}\BibitemShut {NoStop}%
\bibitem [{\citenamefont {Binder}\ \emph
  {et~al.}(2015{\natexlab{a}})\citenamefont {Binder}, \citenamefont
  {Vinjanampathy}, \citenamefont {Modi},\ and\ \citenamefont
  {Goold}}]{Binder15b}%
  \BibitemOpen
  \bibfield  {author} {\bibinfo {author} {\bibfnamefont {F.}~\bibnamefont
  {Binder}}, \bibinfo {author} {\bibfnamefont {S.}~\bibnamefont
  {Vinjanampathy}}, \bibinfo {author} {\bibfnamefont {K.}~\bibnamefont {Modi}},
  \ and\ \bibinfo {author} {\bibfnamefont {J.}~\bibnamefont {Goold}},\ }\href
  {\doibase 10.1088/1367-2630/17/7/075015} {\bibfield  {journal} {\bibinfo
  {journal} {New J. Phys.}\ }\textbf {\bibinfo {volume} {17}},\ \bibinfo
  {pages} {075015} (\bibinfo {year} {2015}{\natexlab{a}})}\BibitemShut
  {NoStop}%
\bibitem [{\citenamefont {Perarnau-Llobet}\ \emph {et~al.}(2015)\citenamefont
  {Perarnau-Llobet}, \citenamefont {Hovhannisyan}, \citenamefont {Huber},
  \citenamefont {Skrzypczyk}, \citenamefont {Tura},\ and\ \citenamefont
  {Ac\'{\i}n}}]{Perarnau-Llobet:15}%
  \BibitemOpen
  \bibfield  {author} {\bibinfo {author} {\bibfnamefont {M.}~\bibnamefont
  {Perarnau-Llobet}}, \bibinfo {author} {\bibfnamefont {K.~V.}\ \bibnamefont
  {Hovhannisyan}}, \bibinfo {author} {\bibfnamefont {M.}~\bibnamefont {Huber}},
  \bibinfo {author} {\bibfnamefont {P.}~\bibnamefont {Skrzypczyk}}, \bibinfo
  {author} {\bibfnamefont {J.}~\bibnamefont {Tura}}, \ and\ \bibinfo {author}
  {\bibfnamefont {A.}~\bibnamefont {Ac\'{\i}n}},\ }\href {\doibase
  10.1103/PhysRevE.92.042147} {\bibfield  {journal} {\bibinfo  {journal} {Phys.
  Rev. E}\ }\textbf {\bibinfo {volume} {92}},\ \bibinfo {pages} {042147}
  (\bibinfo {year} {2015})}\BibitemShut {NoStop}%
\bibitem [{\citenamefont {Binder}(2016)}]{BinderThesis}%
  \BibitemOpen
  \bibfield  {author} {\bibinfo {author} {\bibfnamefont {F.~C.}\ \bibnamefont
  {Binder}},\ }\emph {\bibinfo {title} {Work, heat, and power of quantum
  processes}},\ \href
  {https://ora.ox.ac.uk/objects/uuid:279871ea-3b2e-4baf-975c-1bd42b4961c3}
  {\bibinfo {type} {{DPhil thesis}}},\ \bibinfo  {school} {University of
  Oxford} (\bibinfo {year} {2016})\BibitemShut {NoStop}%
\bibitem [{\citenamefont {Gelbwaser-Klimovsky}\ \emph
  {et~al.}(2015)\citenamefont {Gelbwaser-Klimovsky}, \citenamefont {Niedenzu},\
  and\ \citenamefont {Kurizki}}]{GelbwaserKlimovsky2015329}%
  \BibitemOpen
  \bibfield  {author} {\bibinfo {author} {\bibfnamefont {D.}~\bibnamefont
  {Gelbwaser-Klimovsky}}, \bibinfo {author} {\bibfnamefont {W.}~\bibnamefont
  {Niedenzu}}, \ and\ \bibinfo {author} {\bibfnamefont {G.}~\bibnamefont
  {Kurizki}},\ }in\ \href {\doibase 10.1016/bs.aamop.2015.07.002} {\emph
  {\bibinfo {booktitle} {Advances In Atomic, Molecular, and Optical
  Physics}}},\ Vol.~\bibinfo {volume} {64}\ (\bibinfo  {publisher} {Academic
  Press},\ \bibinfo {year} {2015})\ pp.\ \bibinfo {pages} {329 --
  407}\BibitemShut {NoStop}%
\bibitem [{\citenamefont {Alicki}(1979)}]{Alicki1979}%
  \BibitemOpen
  \bibfield  {author} {\bibinfo {author} {\bibfnamefont {R.}~\bibnamefont
  {Alicki}},\ }\href {\doibase 10.1088/0305-4470/12/5/007} {\bibfield
  {journal} {\bibinfo  {journal} {J. Phys. A: Math. Gen.}\ }\textbf {\bibinfo
  {volume} {12}},\ \bibinfo {pages} {L103} (\bibinfo {year}
  {1979})}\BibitemShut {NoStop}%
\bibitem [{\citenamefont {Deffner}\ and\ \citenamefont
  {Lutz}(2013)}]{Deffner:14}%
  \BibitemOpen
  \bibfield  {author} {\bibinfo {author} {\bibfnamefont {S.}~\bibnamefont
  {Deffner}}\ and\ \bibinfo {author} {\bibfnamefont {E.}~\bibnamefont {Lutz}},\
  }\href {http://stacks.iop.org/1751-8121/46/i=33/a=335302} {\bibfield
  {journal} {\bibinfo  {journal} {J. Phys. A: Math. Th.}\ }\textbf {\bibinfo
  {volume} {46}},\ \bibinfo {pages} {335302} (\bibinfo {year}
  {2013})}\BibitemShut {NoStop}%
\bibitem [{\citenamefont {Wang}\ \emph {et~al.}(2013)\citenamefont {Wang},
  \citenamefont {Vinjanampathy}, \citenamefont {Strauch},\ and\ \citenamefont
  {Jacobs}}]{PhysRevLett.110.157207}%
  \BibitemOpen
  \bibfield  {author} {\bibinfo {author} {\bibfnamefont {X.}~\bibnamefont
  {Wang}}, \bibinfo {author} {\bibfnamefont {S.}~\bibnamefont {Vinjanampathy}},
  \bibinfo {author} {\bibfnamefont {F.~W.}\ \bibnamefont {Strauch}}, \ and\
  \bibinfo {author} {\bibfnamefont {K.}~\bibnamefont {Jacobs}},\ }\href
  {\doibase 10.1103/PhysRevLett.110.157207} {\bibfield  {journal} {\bibinfo
  {journal} {Phys. Rev. Lett.}\ }\textbf {\bibinfo {volume} {110}},\ \bibinfo
  {pages} {157207} (\bibinfo {year} {2013})}\BibitemShut {NoStop}%
\bibitem [{\citenamefont {Hegerfeldt}(2013)}]{PhysRevLett.111.260501}%
  \BibitemOpen
  \bibfield  {author} {\bibinfo {author} {\bibfnamefont {G.~C.}\ \bibnamefont
  {Hegerfeldt}},\ }\href {\doibase 10.1103/PhysRevLett.111.260501} {\bibfield
  {journal} {\bibinfo  {journal} {Phys. Rev. Lett.}\ }\textbf {\bibinfo
  {volume} {111}},\ \bibinfo {pages} {260501} (\bibinfo {year}
  {2013})}\BibitemShut {NoStop}%
\bibitem [{\citenamefont {Hegerfeldt}(2014)}]{PhysRevA.90.032110}%
  \BibitemOpen
  \bibfield  {author} {\bibinfo {author} {\bibfnamefont {G.~C.}\ \bibnamefont
  {Hegerfeldt}},\ }\href {\doibase 10.1103/PhysRevA.90.032110} {\bibfield
  {journal} {\bibinfo  {journal} {Phys. Rev. A}\ }\textbf {\bibinfo {volume}
  {90}},\ \bibinfo {pages} {032110} (\bibinfo {year} {2014})}\BibitemShut
  {NoStop}%
\bibitem [{\citenamefont {Mandelstam}\ and\ \citenamefont
  {Tamm}(1945)}]{Mandelstam:45}%
  \BibitemOpen
  \bibfield  {author} {\bibinfo {author} {\bibfnamefont {L.}~\bibnamefont
  {Mandelstam}}\ and\ \bibinfo {author} {\bibfnamefont {I.}~\bibnamefont
  {Tamm}},\ }\href@noop {} {\bibfield  {journal} {\bibinfo  {journal} {J. Phys.
  (USSR)}\ }\textbf {\bibinfo {volume} {9}},\ \bibinfo {pages} {1} (\bibinfo
  {year} {1945})}\BibitemShut {NoStop}%
\bibitem [{\citenamefont {Margolus}\ and\ \citenamefont
  {Levitin}(1998)}]{Margolus:98}%
  \BibitemOpen
  \bibfield  {author} {\bibinfo {author} {\bibfnamefont {N.}~\bibnamefont
  {Margolus}}\ and\ \bibinfo {author} {\bibfnamefont {L.~B.}\ \bibnamefont
  {Levitin}},\ }\href {\doibase 10.1016/S0167-2789(98)00054-2} {\bibfield
  {journal} {\bibinfo  {journal} {Phys. D.}\ }\textbf {\bibinfo {volume}
  {120}},\ \bibinfo {pages} {188} (\bibinfo {year} {1998})}\BibitemShut
  {NoStop}%
\bibitem [{\citenamefont {Pires}\ \emph {et~al.}(2016)\citenamefont {Pires},
  \citenamefont {Cianciaruso}, \citenamefont {C\'eleri}, \citenamefont
  {Adesso},\ and\ \citenamefont {Soares-Pinto}}]{Pires2016}%
  \BibitemOpen
  \bibfield  {author} {\bibinfo {author} {\bibfnamefont {D.~P.}\ \bibnamefont
  {Pires}}, \bibinfo {author} {\bibfnamefont {M.}~\bibnamefont {Cianciaruso}},
  \bibinfo {author} {\bibfnamefont {L.~C.}\ \bibnamefont {C\'eleri}}, \bibinfo
  {author} {\bibfnamefont {G.}~\bibnamefont {Adesso}}, \ and\ \bibinfo {author}
  {\bibfnamefont {D.~O.}\ \bibnamefont {Soares-Pinto}},\ }\href {\doibase
  10.1103/PhysRevX.6.021031} {\bibfield  {journal} {\bibinfo  {journal} {Phys.
  Rev. X}\ }\textbf {\bibinfo {volume} {6}},\ \bibinfo {pages} {021031}
  (\bibinfo {year} {2016})}\BibitemShut {NoStop}%
\bibitem [{\citenamefont {Binder}\ \emph
  {et~al.}(2015{\natexlab{b}})\citenamefont {Binder}, \citenamefont
  {Vinjanampathy}, \citenamefont {Modi},\ and\ \citenamefont
  {Goold}}]{Binder15a}%
  \BibitemOpen
  \bibfield  {author} {\bibinfo {author} {\bibfnamefont {F.}~\bibnamefont
  {Binder}}, \bibinfo {author} {\bibfnamefont {S.}~\bibnamefont
  {Vinjanampathy}}, \bibinfo {author} {\bibfnamefont {K.}~\bibnamefont {Modi}},
  \ and\ \bibinfo {author} {\bibfnamefont {J.}~\bibnamefont {Goold}},\ }\href
  {\doibase 10.1103/PhysRevE.91.032119} {\bibfield  {journal} {\bibinfo
  {journal} {Phys. Rev. E}\ }\textbf {\bibinfo {volume} {91}},\ \bibinfo
  {pages} {032119} (\bibinfo {year} {2015}{\natexlab{b}})}\BibitemShut
  {NoStop}%
\bibitem [{\citenamefont {Giovannetti}\ \emph
  {et~al.}(2003{\natexlab{a}})\citenamefont {Giovannetti}, \citenamefont
  {Lloyd},\ and\ \citenamefont {Maccone}}]{Giovannetti2003}%
  \BibitemOpen
  \bibfield  {author} {\bibinfo {author} {\bibfnamefont {V.}~\bibnamefont
  {Giovannetti}}, \bibinfo {author} {\bibfnamefont {S.}~\bibnamefont {Lloyd}},
  \ and\ \bibinfo {author} {\bibfnamefont {L.}~\bibnamefont {Maccone}},\ }\href
  {\doibase 10.1103/PhysRevA.67.052109} {\bibfield  {journal} {\bibinfo
  {journal} {Phys. Rev. A}\ }\textbf {\bibinfo {volume} {67}},\ \bibinfo
  {pages} {052109} (\bibinfo {year} {2003}{\natexlab{a}})}\BibitemShut
  {NoStop}%
\bibitem [{\citenamefont {Zander}\ \emph {et~al.}(2007)\citenamefont {Zander},
  \citenamefont {Plastino}, \citenamefont {Plastino},\ and\ \citenamefont
  {Casas}}]{Zander2007}%
  \BibitemOpen
  \bibfield  {author} {\bibinfo {author} {\bibfnamefont {C.}~\bibnamefont
  {Zander}}, \bibinfo {author} {\bibfnamefont {A.~R.}\ \bibnamefont
  {Plastino}}, \bibinfo {author} {\bibfnamefont {A.}~\bibnamefont {Plastino}},
  \ and\ \bibinfo {author} {\bibfnamefont {M.}~\bibnamefont {Casas}},\ }\href
  {\doibase 10.1088/1751-8113/40/11/020} {\bibfield  {journal} {\bibinfo
  {journal} {J. Phys. A: Math. Th.}\ }\textbf {\bibinfo {volume} {40}},\
  \bibinfo {pages} {2861} (\bibinfo {year} {2007})}\BibitemShut {NoStop}%
\bibitem [{\citenamefont {Giovannetti}\ \emph
  {et~al.}(2003{\natexlab{b}})\citenamefont {Giovannetti}, \citenamefont
  {Lloyd},\ and\ \citenamefont {Maccone}}]{Giovannetti2003a}%
  \BibitemOpen
  \bibfield  {author} {\bibinfo {author} {\bibfnamefont {V.}~\bibnamefont
  {Giovannetti}}, \bibinfo {author} {\bibfnamefont {S.}~\bibnamefont {Lloyd}},
  \ and\ \bibinfo {author} {\bibfnamefont {L.}~\bibnamefont {Maccone}},\ }\href
  {\doibase 10.1209/epl/i2003-00418-8} {\bibfield  {journal} {\bibinfo
  {journal} {Europhys. Lett.}\ }\textbf {\bibinfo {volume} {62}},\ \bibinfo
  {pages} {615} (\bibinfo {year} {2003}{\natexlab{b}})}\BibitemShut {NoStop}%
\bibitem [{\citenamefont {Giovannetti}\ \emph
  {et~al.}(2003{\natexlab{c}})\citenamefont {Giovannetti}, \citenamefont
  {Lloyd},\ and\ \citenamefont {Maccone}}]{Giovannetti2003b}%
  \BibitemOpen
  \bibfield  {author} {\bibinfo {author} {\bibfnamefont {V.}~\bibnamefont
  {Giovannetti}}, \bibinfo {author} {\bibfnamefont {S.}~\bibnamefont {Lloyd}},
  \ and\ \bibinfo {author} {\bibfnamefont {L.}~\bibnamefont {Maccone}},\ }in\
  \href {\doibase 10.1117/12.507486} {\emph {\bibinfo {booktitle} {Proc.
  SPIE}}},\ Vol.\ \bibinfo {volume} {5111}\ (\bibinfo {year}
  {2003})\BibitemShut {NoStop}%
\bibitem [{\citenamefont {Xu}(2016)}]{Xu2015}%
  \BibitemOpen
  \bibfield  {author} {\bibinfo {author} {\bibfnamefont {Z.-Y.}\ \bibnamefont
  {Xu}},\ }\href {\doibase 10.1088/1367-2630/18/7/073005} {\bibfield  {journal}
  {\bibinfo  {journal} {New J. Phys.}\ }\textbf {\bibinfo {volume} {18}},\
  \bibinfo {pages} {073005} (\bibinfo {year} {2016})}\BibitemShut {NoStop}%
\bibitem [{\citenamefont {Fr\"owis}(2012)}]{Frowis2012}%
  \BibitemOpen
  \bibfield  {author} {\bibinfo {author} {\bibfnamefont {F.}~\bibnamefont
  {Fr\"owis}},\ }\href {\doibase 10.1103/PhysRevA.85.052127} {\bibfield
  {journal} {\bibinfo  {journal} {Phys. Rev. A}\ }\textbf {\bibinfo {volume}
  {85}},\ \bibinfo {pages} {052127} (\bibinfo {year} {2012})}\BibitemShut
  {NoStop}%
\bibitem [{\citenamefont {Toth}\ and\ \citenamefont
  {Apellaniz}(2014)}]{Toth2014}%
  \BibitemOpen
  \bibfield  {author} {\bibinfo {author} {\bibfnamefont {G.}~\bibnamefont
  {Toth}}\ and\ \bibinfo {author} {\bibfnamefont {I.}~\bibnamefont
  {Apellaniz}},\ }\href {\doibase 10.1088/1751-8113/47/42/424006} {\bibfield
  {journal} {\bibinfo  {journal} {J. Phys. A: Math. Th.}\ }\textbf {\bibinfo
  {volume} {47}},\ \bibinfo {pages} {424006} (\bibinfo {year}
  {2014})}\BibitemShut {NoStop}%
\bibitem [{Note1()}]{Note1}%
  \BibitemOpen
  \bibinfo {note} {Though an exact form for the radius of the separable ball is
  not known, it has been bounded from below and above~\cite {GurvitsBarnum2002,
  GurvitsBarnum2005, AubrunSzarek2006}.}\BibitemShut {Stop}%
\bibitem [{\citenamefont {Giovannetti}\ \emph {et~al.}(2004)\citenamefont
  {Giovannetti}, \citenamefont {Lloyd},\ and\ \citenamefont
  {Maccone}}]{Giovannetti2004}%
  \BibitemOpen
  \bibfield  {author} {\bibinfo {author} {\bibfnamefont {V.}~\bibnamefont
  {Giovannetti}}, \bibinfo {author} {\bibfnamefont {S.}~\bibnamefont {Lloyd}},
  \ and\ \bibinfo {author} {\bibfnamefont {L.}~\bibnamefont {Maccone}},\ }\href
  {\doibase 10.1088/1464-4266/6/8/028} {\bibfield  {journal} {\bibinfo
  {journal} {J. Opt. B: Quant. Sem. Opt.}\ }\textbf {\bibinfo {volume} {6}},\
  \bibinfo {pages} {S807} (\bibinfo {year} {2004})}\BibitemShut {NoStop}%
\bibitem [{\citenamefont {B\"uchler}\ \emph {et~al.}(2007)\citenamefont
  {B\"uchler}, \citenamefont {Micheli},\ and\ \citenamefont {Zoller}}]{zoller}%
  \BibitemOpen
  \bibfield  {author} {\bibinfo {author} {\bibfnamefont {H.~P.}\ \bibnamefont
  {B\"uchler}}, \bibinfo {author} {\bibfnamefont {A.}~\bibnamefont {Micheli}},
  \ and\ \bibinfo {author} {\bibfnamefont {P.}~\bibnamefont {Zoller}},\ }\href
  {http://dx.doi.org/10.1038/nphys678} {\bibfield  {journal} {\bibinfo
  {journal} {Nat. Phys.}\ }\textbf {\bibinfo {volume} {3}},\ \bibinfo {pages}
  {726} (\bibinfo {year} {2007})}\BibitemShut {NoStop}%
\bibitem [{\citenamefont {Boixo}\ \emph {et~al.}(2007)\citenamefont {Boixo},
  \citenamefont {Flammia}, \citenamefont {Caves},\ and\ \citenamefont
  {Geremia}}]{Boxio:07}%
  \BibitemOpen
  \bibfield  {author} {\bibinfo {author} {\bibfnamefont {S.}~\bibnamefont
  {Boixo}}, \bibinfo {author} {\bibfnamefont {S.~T.}\ \bibnamefont {Flammia}},
  \bibinfo {author} {\bibfnamefont {C.~M.}\ \bibnamefont {Caves}}, \ and\
  \bibinfo {author} {\bibfnamefont {J.}~\bibnamefont {Geremia}},\ }\href
  {\doibase 10.1103/PhysRevLett.98.090401} {\bibfield  {journal} {\bibinfo
  {journal} {Phys. Rev. Lett.}\ }\textbf {\bibinfo {volume} {98}},\ \bibinfo
  {pages} {090401} (\bibinfo {year} {2007})}\BibitemShut {NoStop}%
\bibitem [{\citenamefont {Napolitano}\ \emph {et~al.}(2011)\citenamefont
  {Napolitano}, \citenamefont {Koschorreck}, \citenamefont {Dubost},
  \citenamefont {Behbood}, \citenamefont {Sewell},\ and\ \citenamefont
  {Mitchell}}]{napolitano2011interaction}%
  \BibitemOpen
  \bibfield  {author} {\bibinfo {author} {\bibfnamefont {M.}~\bibnamefont
  {Napolitano}}, \bibinfo {author} {\bibfnamefont {M.}~\bibnamefont
  {Koschorreck}}, \bibinfo {author} {\bibfnamefont {B.}~\bibnamefont {Dubost}},
  \bibinfo {author} {\bibfnamefont {N.}~\bibnamefont {Behbood}}, \bibinfo
  {author} {\bibfnamefont {R.}~\bibnamefont {Sewell}}, \ and\ \bibinfo {author}
  {\bibfnamefont {M.~W.}\ \bibnamefont {Mitchell}},\ }\href {\doibase
  10.1038/nature09778} {\bibfield  {journal} {\bibinfo  {journal} {Nature}\
  }\textbf {\bibinfo {volume} {471}},\ \bibinfo {pages} {486} (\bibinfo {year}
  {2011})}\BibitemShut {NoStop}%
\bibitem [{\citenamefont {Dicke}(1954)}]{PhysRev.93.99}%
  \BibitemOpen
  \bibfield  {author} {\bibinfo {author} {\bibfnamefont {R.~H.}\ \bibnamefont
  {Dicke}},\ }\href {\doibase 10.1103/PhysRev.93.99} {\bibfield  {journal}
  {\bibinfo  {journal} {Phys. Rev.}\ }\textbf {\bibinfo {volume} {93}},\
  \bibinfo {pages} {99} (\bibinfo {year} {1954})}\BibitemShut {NoStop}%
\bibitem [{\citenamefont {Lipkin}\ \emph {et~al.}(1965)\citenamefont {Lipkin},
  \citenamefont {Meshkov},\ and\ \citenamefont {Glick}}]{LIPKIN1965188}%
  \BibitemOpen
  \bibfield  {author} {\bibinfo {author} {\bibfnamefont {H.}~\bibnamefont
  {Lipkin}}, \bibinfo {author} {\bibfnamefont {N.}~\bibnamefont {Meshkov}}, \
  and\ \bibinfo {author} {\bibfnamefont {A.}~\bibnamefont {Glick}},\ }\href
  {\doibase 10.1016/0029-5582(65)90862-X} {\bibfield  {journal} {\bibinfo
  {journal} {Nuclear Physics}\ }\textbf {\bibinfo {volume} {62}},\ \bibinfo
  {pages} {188 } (\bibinfo {year} {1965})}\BibitemShut {NoStop}%
\bibitem [{\citenamefont {M\o{}lmer}\ and\ \citenamefont
  {S\o{}rensen}(1999)}]{MolmerSorensen1998}%
  \BibitemOpen
  \bibfield  {author} {\bibinfo {author} {\bibfnamefont {K.}~\bibnamefont
  {M\o{}lmer}}\ and\ \bibinfo {author} {\bibfnamefont {A.}~\bibnamefont
  {S\o{}rensen}},\ }\href {\doibase 10.1103/PhysRevLett.82.1835} {\bibfield
  {journal} {\bibinfo  {journal} {Phys. Rev. Lett.}\ }\textbf {\bibinfo
  {volume} {82}},\ \bibinfo {pages} {1835} (\bibinfo {year}
  {1999})}\BibitemShut {NoStop}%
\bibitem [{\citenamefont {Modi}\ \emph {et~al.}(2011)\citenamefont {Modi},
  \citenamefont {Cable}, \citenamefont {Williamson},\ and\ \citenamefont
  {Vedral}}]{prx}%
  \BibitemOpen
  \bibfield  {author} {\bibinfo {author} {\bibfnamefont {K.}~\bibnamefont
  {Modi}}, \bibinfo {author} {\bibfnamefont {H.}~\bibnamefont {Cable}},
  \bibinfo {author} {\bibfnamefont {M.}~\bibnamefont {Williamson}}, \ and\
  \bibinfo {author} {\bibfnamefont {V.}~\bibnamefont {Vedral}},\ }\href
  {\doibase 10.1103/PhysRevX.1.021022} {\bibfield  {journal} {\bibinfo
  {journal} {Phys. Rev. X}\ }\textbf {\bibinfo {volume} {1}},\ \bibinfo {pages}
  {021022} (\bibinfo {year} {2011})}\BibitemShut {NoStop}%
\bibitem [{\citenamefont {Gurvits}\ and\ \citenamefont
  {Barnum}(2002)}]{GurvitsBarnum2002}%
  \BibitemOpen
  \bibfield  {author} {\bibinfo {author} {\bibfnamefont {L.}~\bibnamefont
  {Gurvits}}\ and\ \bibinfo {author} {\bibfnamefont {H.}~\bibnamefont
  {Barnum}},\ }\href {\doibase 10.1103/PhysRevA.66.062311} {\bibfield
  {journal} {\bibinfo  {journal} {Phys. Rev. A}\ }\textbf {\bibinfo {volume}
  {66}},\ \bibinfo {pages} {062311} (\bibinfo {year} {2002})}\BibitemShut
  {NoStop}%
\bibitem [{\citenamefont {Gurvits}\ and\ \citenamefont
  {Barnum}(2005)}]{GurvitsBarnum2005}%
  \BibitemOpen
  \bibfield  {author} {\bibinfo {author} {\bibfnamefont {L.}~\bibnamefont
  {Gurvits}}\ and\ \bibinfo {author} {\bibfnamefont {H.}~\bibnamefont
  {Barnum}},\ }\href {\doibase 10.1103/PhysRevA.72.032322} {\bibfield
  {journal} {\bibinfo  {journal} {Phys. Rev. A}\ }\textbf {\bibinfo {volume}
  {72}},\ \bibinfo {pages} {032322} (\bibinfo {year} {2005})}\BibitemShut
  {NoStop}%
\bibitem [{\citenamefont {Aubrun}\ and\ \citenamefont
  {Szarek}(2006)}]{AubrunSzarek2006}%
  \BibitemOpen
  \bibfield  {author} {\bibinfo {author} {\bibfnamefont {G.}~\bibnamefont
  {Aubrun}}\ and\ \bibinfo {author} {\bibfnamefont {S.~J.}\ \bibnamefont
  {Szarek}},\ }\href {\doibase 10.1103/PhysRevA.73.022109} {\bibfield
  {journal} {\bibinfo  {journal} {Phys. Rev. A}\ }\textbf {\bibinfo {volume}
  {73}},\ \bibinfo {pages} {022109} (\bibinfo {year} {2006})}\BibitemShut
  {NoStop}%
\bibitem [{\citenamefont {Poulin}\ \emph {et~al.}(2011)\citenamefont {Poulin},
  \citenamefont {Qarry}, \citenamefont {Somma},\ and\ \citenamefont
  {Verstraete}}]{Poulin2011}%
  \BibitemOpen
  \bibfield  {author} {\bibinfo {author} {\bibfnamefont {D.}~\bibnamefont
  {Poulin}}, \bibinfo {author} {\bibfnamefont {A.}~\bibnamefont {Qarry}},
  \bibinfo {author} {\bibfnamefont {R.}~\bibnamefont {Somma}}, \ and\ \bibinfo
  {author} {\bibfnamefont {F.}~\bibnamefont {Verstraete}},\ }\href {\doibase
  10.1103/PhysRevLett.106.170501} {\bibfield  {journal} {\bibinfo  {journal}
  {Phys. Rev. Lett.}\ }\textbf {\bibinfo {volume} {106}},\ \bibinfo {pages}
  {170501} (\bibinfo {year} {2011})}\BibitemShut {NoStop}%
\end{thebibliography}%
\clearpage
\newpage
\clearpage
\appendix
\section*{Supplementary Material: Enhancing the
charging power of quantum batteries}
\section{Relation between constraints}
\label{s:bounds}

\noindent
Here we show that $\Delta E_\sharp \le \mathcal{E}_\sharp$ and $E_\sharp \le 2\mathcal{E}_\sharp $ by direct computation. In the first case we have
\begin{align}
    \Delta E_\sharp & = \frac{1}{T_\sharp}\int_0^{T_\sharp} dt \sqrt{\tr[ \bm{H}^2\bm{\rho}(t)] - \tr[ \bm{H}\bm{\rho}(t)]^2 }, \notag\\
    & \leq \frac{1}{T_\sharp}\int_0^{T_\sharp} dt \sqrt{\tr[ \bm{H}^2\bm{\rho}(t)]} \notag\\
    & \leq \frac{1}{T_\sharp}\int_0^{T_\sharp} dt \, \lVert \bm{H} \rVert_{\rm op} = \mathcal{E}_\sharp,
\end{align}
where we have used $\tr[ \bm{H}^2\bm{\rho}(t)] \le \lVert \bm{H}^2 \rVert_{\rm op} = \lVert \bm{H} \rVert_{\rm op}^2$ to get to the final line.

Similarly, the time-averaged energy
\begin{align}
    E_\sharp & = \frac{1}{T_\sharp}\int_0^{T_\sharp} dt \,  \{\tr[ \bm{H}\bm{\rho}(t)] - \bm{h}_g\} , \notag\\
    & \leq \frac{1}{T_\sharp}\int_0^{T_\sharp} dt \,  \{\lVert  \bm{H} \rVert_{\rm op} + |\bm{h}_g| \} , \notag\\
    & \leq \frac{2}{T_\sharp}\int_0^{T_\sharp} dt \, \lVert  \bm{H} \rVert_{\rm op}
     = 2\mathcal{E}_\sharp. 
\end{align}
Thus, if the time-averaged operator norm of the Hamiltonian is bounded,  $\Delta E_\sharp$ and $E_\sharp$ are also bounded.

\section{Proof of Theorem~\ref{thm:circuit}}
\label{s:th2}

\noindent{\it Proof.} 
Consider $W(t)=\tr[\bm{I}\bm{\rho}(t)]-\tr[\bm{I}\rho^{\otimes N}]$, the average work done on the system up to time $t$ during the charging process. The instantaneous power is given by $P(t)=d_t W(t) = i \tr\{[\bm{H} ,\bm{I}] \bm{\rho}(t)\}$. The strict inequality
\begin{gather}
    \label{eq:inequality_power}
    P_\sharp = \frac{1}{T_{\sharp}}\int_0^{T_{\sharp}} dt P(t) < \max_{\bm{H}} \left\{ \lVert [\bm{H},\bm{I}] \rVert_{\rm op} \right\} =: P^{\uparrow}
\end{gather}
follows from the fact that any unitary charging has to have vanishing instantaneous power for times $t=0$ and $t={T_\sharp}$. We now evaluate the commutator $[\bm{H},\bm{I}]$ in order to find an upper bound $P^{\uparrow}$ for the average power $P_\sharp$, remembering that we have $\bm{I}=\sum_j I^{(j)}$ and $\bm{H} = \sum_{\mu=1}^s h_{\mu}$. We will use the subscript $\bar{\mu}$ to indicate the set of battery indices that are not included in partition defined by $\mu$. Using the commutation relation between $h_{\mu}$ and $I^{(j)}$ we obtain
\begin{align}
   [\bm{H},\bm{I}] &= \sum_{\mu=1}^s  \left[ h_{\mu} \otimes \mathbb{1}_{\bar{\mu}} \,, \textstyle{\sum_{j=1}^k} I^{(\mu_j)} \otimes \mathbb{1}_{\bar{\mu_j}} \right] \nonumber \\
   &=\sum_\mu  \left[ h_\mu \,, \textstyle{\sum_{j=1}^k} I^{(\mu_j)} \otimes \mathbb{1}_{i\neq \mu_j \in \mu} \right] \otimes \mathbb{1}_{\bar{\mu}}.
\end{align}

Using the definition for $\bm{I}$, it
follows from direct calculation that $\lVert \textstyle{\sum_{i=1}^k} I^{(\mu_i)} \rVert_{\rm op} = \lambda_d k$. Let us define $\alpha_\mu := \lVert h_\mu  \rVert_{\rm op}$ and introduce two normalized operators $X_\mu$ and $\iota_\mu$, as follows:
\begin{gather}
    X_\mu = \frac{h_\mu}{\alpha_\mu},
\quad    
    \iota_{\mu} = \frac{1}{\lambda_d k}{\sum_{i=1}^k I^{(\mu_i)}}.
\end{gather}
Using these, we can rewrite the commutator as
\begin{align}
   [\bm{H},\bm{I}] &= 2 \cdot \lambda_d k \sum_\mu \alpha_\mu\frac{1}{2} [X_\mu,\iota_{\mu}]\otimes\mathbb{1}_{\bar{\mu}} \nonumber \\
   \label{eq:power_y}
    & = 2\lambda_d k \sum_\mu \alpha_\mu Y_\mu  \otimes \mathbb{1}_{\bar{\mu}},
\end{align}
where $Y_\mu = \frac{1}{2} [X_\mu,\iota_{\mu}]$, such that $\lVert Y_\mu \rVert_{\rm op} \leq 1$.

At any time $t$, the operator norm of the Hamiltonian $\bm{H}$ is given by
\begin{gather}
    \label{eq:piecewise_norm}
    \lVert \bm{H} \rVert_{\rm op}  =   \bigg\lVert \sum_{\mu=1}^s h_\mu \bigg\rVert_{\rm op} 
     = \sum_{\mu=1}^s \lVert h_\mu \rVert_{\rm op}
     = \sum_{\mu=1}^s |\alpha_\mu |,
\end{gather}
where the equality holds due to the fact that, at each step in time, every term $h_\mu$ acts on a different $k$-partition of the Hilbert space. 
Accordingly, we obtain that
\begin{gather}
\label{eq:constraint_again}
    \frac{1}{T_{\sharp}} \int_0^{T_{\sharp}} dt \lVert \bm{H} \rVert_{\rm op}  =  \frac{1}{T_{\sharp}} \int_0^{T_{\sharp}} dt \sum_{\mu=1}^s |\alpha_\mu | \leq N\mathcal{E}.
\end{gather}

Now we consider the expression given in Eq.~\eqref{eq:power_y}, to calculate the upper bound $P^{\uparrow}$. Once again, we use the fact that at each step in time there are $s$ terms acting on different $k$-partitions of the Hilbert space, such that the operator norm of $[\bm{H},\bm{I}]$ can be calculated exactly,
\begin{align}
    \lVert [\bm{H},\bm{I}] \rVert_{\rm op} & = 2\lambda_d k  \bigg\lVert \sum_{\mu=1}^s \alpha_\mu Y_\mu \otimes \mathbb{1}_{\bar{\mu}} \bigg\rVert_{\rm op} \nonumber\\
    & = 2\lambda_d k  \sum_{\mu=1}^s \lVert \alpha_\mu Y_\mu \otimes \mathbb{1}_{\bar{\mu}} \rVert_{\rm op} \nonumber\\
    \label{eq:piecewise_norm_y}
    & \leq 2\lambda_d k  \sum_{\mu=1}^s |\alpha_\mu |,
\end{align}
where the inequality in line Eq.~\eqref{eq:piecewise_norm_y} holds due to the fact that $\lVert Y_\mu \rVert_{\rm op} \leq 1$ by definition, and where the sum can be carried out of the operator norm thanks to the fact that, at any given time, the $s$ subgroups of $k$ batteries are not overlapping.
Using Eq.~\eqref{eq:constraint_again}, we obtain
\begin{align}
    \frac{1}{T_{\sharp}} \int_0^{T_{\sharp}} dt \lVert  [\bm{H},\bm{I}]  \rVert_{\rm op}  &= 2\lambda_d k  \frac{1}{T_{\sharp}} \int_0^{T_{\sharp}} dt \sum_{\mu=1}^s |\alpha_\mu | \nonumber \\
    &\leq 2\lambda_d k N\mathcal{E}.
\end{align}
Plugging this result back into Eq.~\eqref{eq:inequality_power}, we get 
\begin{align}
& \frac{1}{T_{\sharp}}\int_0^{T_{\sharp}} dt P(t)  < P^{\uparrow} 
= 2\lambda_d k N \mathcal{E}.
\end{align}
We calculate the quantum advantage as in Eq.~\eqref{qadvantage}, where  $P_\|$ is given by the ratio between $W_\|$ and $T_\|=\beta \mathcal{L}_1/\min\{E,\Delta E\}$. 
Work $W_\|=N W$ is extensive, and in general $W=q 2\lambda_d$, where $0< q \leq 1$.
Thus, we obtain
\begin{align}
   \Gamma_{\rm C0}<\frac{2\lambda_d k N \mathcal{E}}{2\lambda_d N q \frac{\min\{E,\Delta E\}}{\beta\mathcal{L}_1}} =  \frac{\beta \mathcal{L}_1 \mathcal{E}}{q \: \min\{E,\Delta E\}} k,
\end{align}
as we intended to prove. \hfill $\blacksquare$

\begin{figure}[t]
\centering
\includegraphics[width=0.42\textwidth]
{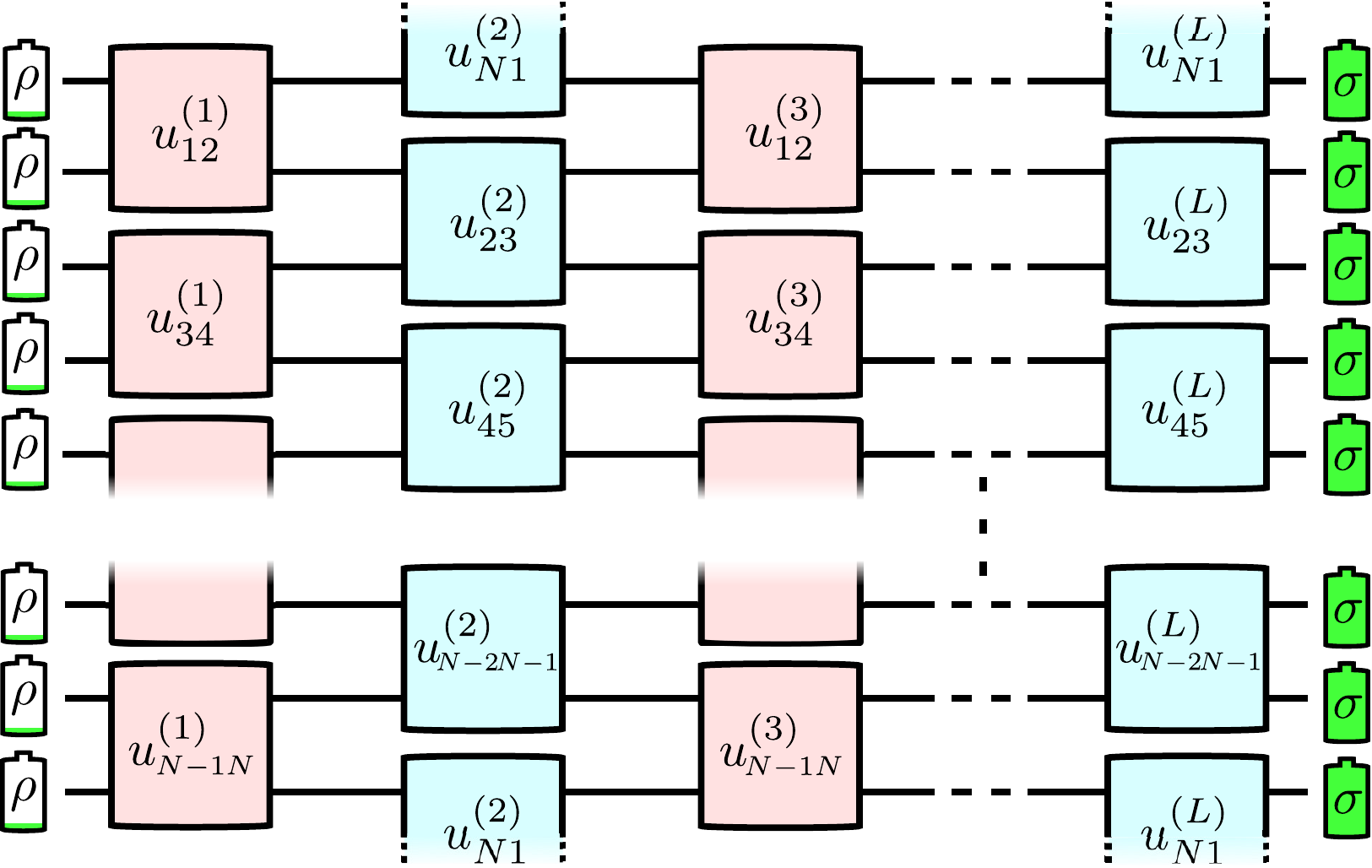}
\caption{\textbf{(Color online) Unitary circuit with k=2.} At each time step $t$, the driving Hamiltonian consists of a set of $s=N/2$ terms, each of which involving a different pair of batteries $i,j$. The result is $s$ independent unitary operations $u_{ii+1}^{(t)}$ acting on pairs $i,i+1$. Note that at each step, there is no overlap between different pairs of batteries; at every successive step the pairs are changed in order to allow the formation of highly entangled states. A circuit of this type can be used to approximate any time-dependent unitary evolution $\bm{U}(t)$~\cite{Poulin2011}, with precision that increases with the number of steps $L$. The implementation of $\bm{U}$ requires an extra amount of time that depends on the number of non-commuting terms in the Hamiltonian.}
\label{fig:pairwise_circuit}
\end{figure}

\section{Proof of Theorem~\ref{thm:reach}}
\label{s:co1}
\noindent{\it Proof.}
Here our goal is to relate a generic unitary evolution to a circuit based charging procedure. We first note that the charging Hamiltonian can always be decomposed into a number $M$ of non-commuting terms:
\begin{gather}
\label{eq:neighbour}
   \bm{H}(t) =\sum_{j=1}^M \bm{H}^{(j)}(t) 
   \quad \mbox{with} \quad
   \bm{H}^{(j)}(t) =\sum_{\mu=1}^s h^{(j)}_{\mu}(t),
\end{gather}
where $[h^{(j)}_{\mu}(t),h^{(j)}_{\mu'}(t)]=0$; this decomposition is in general different for different values of $t$.
The unitary evolution $\bm{U}$ generated by this time-dependent Hamiltonian can always be approximated, using the Trotter-Suzuki decomposition~\cite{Poulin2011}, by the following product of unitary tranformations:
\begin{gather}
\label{eq:trotter}
    \bm{U}_{\rm Trot}= \prod_{l=1}^L \prod_{j=1}^M\exp\left[ - i \bm{H}^{(j)}\left(\frac{lT_\sharp}{L}\right) \frac{T_\sharp}{L} \right].
\end{gather}
In the limit $L\rightarrow\infty$, $\bm{U}_{\rm Trot} = \bm{U}$; however, they do not correspond to the same implementation: The Hamiltonian generating $\bm{U}_{\rm Trot}$ is piecewise time-independent and in the circuit form discussed in Sec.~\ref{s:th2}. Since each of the $M$ terms at each time step must be implemented sequentially, $\bm{U}_{\rm Trot}$ takes $M$ times longer to run than $\bm{U}$, with a corresponding drop in power  $P_\sharp= M P_{\textrm{Trot}}$.

Since we have an upper bound, from Theorem~\ref{thm:circuit}, on the quantum advantage for circuit model Hamiltonians, and the power for a more general Hamiltonian is at most $M$ times greater, it must be that $\Gamma_{\rm C0} < M\gamma k$ in this case. In order to complete the proof, we now need to consider how the minimum necessary value of $M$ scales with $k$ and $m$.

The quantity $m$ denotes the maximum number of other batteries any one can interact with. In order for the number of terms $M$ in Eq.~\eqref{eq:neighbour} to be sufficient for the required decomposition, it must at least equal the largest possible number of different $k$-partitions $\mu$ that have a non-trivial amount of indices in common, while containing the same index $\mu_i$ at most $m$ times. Let us provide a few simple examples to clarify the meaning of $M$, where we will assume that $N$ can be arbitrarily large.

\paragraph*{$(k=2, m=1)$} In this case $M$ is trivially equal to 1. A possible choice is given by the first 2-partition $(1,2)$, after which any other partition $(i,j)$ can contain neither 1 nor 2. This has to be true for any choice of other partitions, therefore $M=1$. In other words, in this case, the trotterization is not necessary and the unitary can be perfectly simulated with a piecewise unitary circuit. 

\paragraph*{$(k=2, m=2)$} Let us start with the first 2-partition $(1,2)$, followed by $(2,3)$ and $(1,3)$. Any other choice of two indices would form a partition that does not contain any element of at least one of the previous three, thus $M=3$. In this case the simulating circuit is at most 3 times slower than the actual unitary.

\paragraph*{$(k=3, m=2)$} Now the first 3-partition $(1,2,3)$ is followed by $(1,4,5)$, $(2,4,6)$ and $(3,5,6)$. Any other choice of three indices would form a partition that does not contain any element of at least one of the previous four, thus $M=4$.

In general, for a given $k$ and a given $m$, we could start -- without loss of generality -- from the first \emph{ordered} partition $(1,\dots,k)$. Remembering that each of those indices can appear at most $m$ times, we can construct $m$ sets containing $1$, followed by $m-1$ sets containing $2$, $3$, $4$ and so on until $m-1$ sets containing $k$, for a total of $k(m-1)+1$ terms. In the worst case scenario, all of these partitions have at least one element in common. However, any subsequent partition cannot contain any of the indices included in the first ordered partition $(1,\dots,k)$, thus $M\leq k(m-1)+1$.

Taking this most general, worst case scenario, we have a bound on the quantum advantage given by
\begin{gather}
\label{eq:conservative_bound_final}
    \Gamma_{\rm C0} <  (k(m-1)+1)\gamma k
    = \gamma \left( k^2 (m-1) + k\right),
\end{gather}
where $\gamma := \frac{\beta \mathcal{L}_1 \mathcal{E}}{q \min\{E,\Delta E\}}$. \hfill $\blacksquare$
\vspace{5pt}
\section{On Conjecture~\ref{conj:conj}.}
\label{s:conjecture}

\noindent
Let us consider a general time-dependent Hamiltonian that contains all the possible $k$-body interaction terms between the $N$ batteries that constitute the system, \textit{i.e.}, $\bm{H} = \sum_\mu h_\mu$ contains $N!/k!(N-k)!$ terms in the sum. With the aim of obtaining an upper bound for the quantum advantage under the constraint C0, we follow the proof provided for Theorem 2, until Eq.~\eqref{eq:power_y}. We then find an explicit relation between the elements of $X_\mu$ and those of $Y_\mu$. Let us consider the product basis $\mathcal{B}_\mu:=\{|a\rangle_\mu\}$ for the subset of batteries defined by $\mu$. Each element $|a\rangle_\mu = \otimes_{i=1}^k |a_i\rangle_{\mu_i}$ is a product state of the partition of the Hilbert space associated with $\mu$. In this basis we can write
\begin{gather}
    \iota_{\mu} = \sum_a \eta_a |a\rangle
    _\mu\langle a|, \\
    \label{eq:x}
    X_\mu = \sum_{a<b} \Big(x_{ab}^\mu |a\rangle_\mu\langle b| + h.c.\Big) + \sum_{a}x^{\mu}_{aa} |a\rangle_\mu\langle a|.
\end{gather}
By explicit calculation using this basis we obtain
\begin{gather}
\label{eq:y}
    Y_\mu = \sum_{a<b} \bigg(\frac{\eta_b-\eta_a}{2}\bigg) \Big( x_{a,b}^\mu |a\rangle_\mu\langle b| - h.c. \Big),
\end{gather}
where $|\eta_a|<1$ and $0<(\eta_b - \eta_a)/2\leq 1$ for $a<b$, due to the structure of $\iota_{\mu}$. Our conjecture reduces to the following: 
\begin{gather}
\label{eq:conjecture}
  \lVert\textstyle{\sum_\mu}  \alpha_\mu Y_\mu  \otimes \mathbb{1}_{\bar{\mu}} \rVert_{\rm op} \leq 
    \lVert \textstyle{\sum_\mu} \alpha_\mu X_\mu  \otimes \mathbb{1}_{\bar{\mu}} \rVert_{\rm op},
\end{gather}
which is itself upper bounded by $N\mathcal{E}$. If Eq.~\eqref{eq:conjecture} holds, then for any choice of time-dependent $k$-body interaction Hamiltonian $\bm{H}$, subject to constraint C0, the average power $P_\sharp$ is upper bounded by $2\lambda_d k N \mathcal{E}$, thus, $\Gamma_{\rm C0} < \beta \mathcal{L}_1 k$. 

An extensive numerical search failed to find any counterexamples to our conjecture.
We have calculated the quantity $P=\lVert \textstyle\sum_\mu  \alpha_\mu Y_\mu  \otimes \mathbb{1}_{\bar{\mu}} \rVert_{\rm op}/\lVert \textstyle\sum_\mu  \alpha_\mu X_\mu  \otimes \mathbb{1}_{\bar{\mu}} \rVert_{\rm op}$ for a large set of charging Hamiltonians, and found it to always be smaller than the unit, as conjectured. In order to generate these Hamiltonians, we sample unitaries $u_\mu$ according to the Haar measure, and obtain $h_\mu = i \log[u_\mu]$, where $\log[A]$ is the natural matrix logarithm of $A$; $P$ is calculated explicitly for every sampled Hamiltonian. With a sample size $\nu = 10^5$ we ran the simulation for $(N,k)$ equal to $(3,2)$, $(4,2)$, $(4,3)$ and $(6,2)$. Not a single instance of $P>1$ was recorded.
This numerical evidence does not represent a proof of our conjecture since there could be a measure zero set of Hamiltonians for which $P>1$.


\end{document}